\documentclass[11pt]{article}
\usepackage {amsfonts}
\usepackage {graphicx}
\usepackage {epsfig}
\usepackage{amssymb}
\usepackage{amsmath}
\usepackage {longtable}
\usepackage[english]{babel}
\begin{document}

\title{Spin rotation and birefringence effect for a particle in a
high energy storage ring and measurement of the real part of the
coherent elastic zero-angle scattering amplitude, electric and
magnetic polarizabilities}

\author{V.G. Baryshevsky, A.A. Gurinovich \\ Research Institute for Nuclear Problems, Belarusian State
University,\\ 11 Bobruyskaya Str., Minsk 220050, Belarus,
\\ e-mail: bar@inp.minsk.by}

\maketitle

\begin{center}
\begin{abstract}
In the present paper the equations for the spin evolution of a
particle in a storage ring are analyzed considering contributions
from the tensor electric and magnetic polarizabilities of the
particle. Study of spin rotation and birefringence effect for a
particle in a high energy storage ring provides for measurement as
the real part of the coherent elastic zero-angle scattering
amplitude as well as tensor electric and magnetic
polarizabilities.

We proposed the method for measurement the real part of the
elastic coherent zero-angle scattering amplitude of particles and
nuclei in a storage ring by the paramagnetic resonance in the
periodical in time nuclear pseudoelectric and pseudomagnetic
fields.
\end{abstract}
\end{center}
%\maketitle

\section{INTRODUCTION}
% from proposal - take the rfeerencies

Investigation of spin-dependent interactions of elementary
particles at high energies is a very important part of program of
scientific research has been preparing for carry out at storage
rings (RHIC, CERN, COSY, GSI). It is well known in experimental
particle physics how to measure  a total spin-dependent
cross-section of proton-proton (pp) and proton-deuteron (pd) or
proton-nucleus (pN) and deuteron-nucleus (dN) interactions.

Through analicity we can get dispersion relations between the real
and imaginary parts of the forward scattering amplitude. These
relations are very valuable for analyzing interactions, especially
if we know both real and imaginary parts of the forward scattering
amplitude in a broad energy range through independent experimental
measurements.

There are several experimental possibilities for the indirect
measurement of the real part of the forward scattering amplitude
\cite{Lehar}.

Since no scattering experiment is possible in the forward
direction, the determination of the real part of the forward
amplitudes has always consisted in the measurement of well chosen
elastic scattering observables at small angles and then in the
extrapolation of these observables towards zero angle
\cite{Lehar}. All of these methods, however, contain discrete
ambiguities in the reconstruction of the forward scattering
matrix, which can be removed only by new independent measurements.
Consequently,what is needed is a direct reconstruction of the real
part of the forward scattering matrix such we have in the case of
the imaginary part through the measurement of a total cross
section.

It has been shown in
%[2-9]
\cite{2rot}-\cite{8rot} that there is an unambiguous method which
makes the direct measurement of the real part of the
spin-dependent forward scattering amplitude in the high energy
range possible. This technique is based on the effect proton
(deuteron, antiproton) beam spin rotation in a polarized nuclear
target and on the phenomenon of deuteron spin rotation and
oscillation in a nonpolarized target. This technique uses the
measurement of angle of spin rotation of high energy proton
(deuteron, antiproton) in conditions of transmission experiment -
the so-called spin rotation experiment.

The analogous phenomenon for thermal neutrons was theoretically
predicted in \cite{1rot} and experimentally observed in
\cite{9rot}-\cite{11} (the phenomena of nuclear precession of
neutron spin in a nuclear pseudomagnetic field of a target).

Spin rotation and oscillation experiments as well as investigation
of spin dichroism (i.e. investigation of dependence of beam
absorption on spin orientation) also allow to carry out new
experiments to study P- and T-odd interactions \cite{PT}. Deuteron
spin rotation and oscillation experiments allow to measure the
tensor electric polarizability and, as it is shown below, the
tensor magnetic polarizability, too.
 Change of spin state of a
particle at passing deep into target can influence the
experiments, studying nonelastic processes at collisions of
polarized nucleons and nuclei.
 This impels to investigate possible
influence of spin rotation on the cross-section of such processes.
%

%%%% from birefringence
{Observation of particle spin rotation and birefringence effect
with a storage ring requires to cancel influence of $(g-2)$
precession ($g$ is the gyromagnetic ratio).
 This precession appears due to interaction of the particle magnetic moment
 with an external electromagnetic field.
 The requirement for $(g-2)$ precession influence cancellation also arises when searching for a deuteron electric
dipole moment (EDM)  by the deuteron spin precession in an
electric field in a storage ring \cite {6, project}.
%%%

In the present paper it is shown that the influence of particle
spin precession in a magnetic field on the process of measurement
of spin rotation of a particle passed through a polarized target
can be eliminated with the aid of making the vector (tensor)
polarization of the target rotating (oscillating) with the
frequency coinciding with the frequency of particle spin rotation
due to the particle magnetic moment interacting with a magnetic
field i.e. providing for the paramagnetic resonance under the
action of periodic in time pseudomagnetic (pseudoelectric) field
of the target.

\section{Interactions contributing to the spin motion of a particle in a storage ring}

Considering evolution of the spin of a particle in a storage ring
one should take into account several interactions:

1. interactions of the magnetic and electric dipole moments with
an electromagnetic field;

2. interaction of the particle with the  electric field due to the
tensor electric polarizability;

3. interaction of the particle with the magnetic field due to the
tensor magnetic polarizability;

4. interaction of the particle with the pseudoelectric and
pseudomagnetic nuclear fields of matter.

The equation for the particle spin wavefunction considering all
these interactions is as follows:
\begin{equation}
i\hbar\frac{\partial\Psi(t)}{\partial
t}=\left(\hat{H}_{0}+\hat{V}_{EDM}+\hat{V}_{\vec{E}}+\hat{V}_{\vec{B}}+\hat{V}_E^{nucl}+
\hat{V}_{{B}}^{nucl}\right)\Psi(t) \label{1}
\end{equation}
where $\Psi(t)$ is the particle spin wavefunction,

{$\hat{H}_{0}$ is the Hamiltonian describing the spin behavior
caused by interaction of the magnetic moment with the
electromagnetic field (equation (\ref{1}) with the only
$\hat{H}_{0}$  summand converts to the Bargman-Myshel-Telegdy
equation),}

$\hat{V}_{EDM}$ describes interaction of the particle EDM with the
electric field,
\begin{eqnarray}
\hat{V}_{EDM} & = &
-d\left(\vec{\beta}\times\vec{B}+\vec{E}\right)\vec{S},
%\eqno(2.10)
\label{VEDM}
\end{eqnarray}

$\hat{V}_{\vec{E}}$ describes interaction of the particle with the
electric field due to the tensor electric polarizability:
\begin{equation}
\hat{V}_{\vec{E}}=-\frac{1}{2}\hat{\alpha}_{ik}(E_{eff})_{i}(E_{eff})_{k},
 \label{VE}
\end{equation}
where $\hat{\alpha}_{ik}$ is the electric polarizability tensor of
the particle  , $\vec{E}_{eff}=(\vec{E}+\vec{\beta} \times
\vec{B})$ is the effective electric field; the expression
(\ref{VE}) can be rewritten as follows:
\begin{eqnarray}
\hat{V}_{\vec{E}} =
\alpha_{S}E^{2}_{eff}-\alpha_{T}E^{2}_{eff}\left(\vec{S}\vec{n}_{E}\right)^{2},~
\vec{n}_{E}  =
\frac{\vec{E}+\vec{\beta}\times\vec{B}}{|\vec{E}+\vec{\beta}\times\vec{B}|}
 \label{VE1}
\end{eqnarray}
where $\alpha_{S}$ is the scalar electric polarizability and
$\alpha_{T}$ is the tensor electric polarizability of the
particle.

A particle with the spin $S \ge 1$ also has the magnetic
polarizability which is described by the magnetic polarizability
tensor $\hat{\beta}_{ik}$ and interaction of the particle with the
magnetic field due to the tensor magnetic polarizability is as
follows:
\begin{equation}
\hat{V}_{\vec{B}}=-\frac{1}{2}\hat{\beta}_{ik}(B_{eff})_{i}(B_{eff})_{k},
 \label{VB}
\end{equation}
where $(B_{eff})_{i}$ are the components of the effective magnetic
field $\vec{B}_{eff}=(\vec{B}-\vec{\beta} \times \vec{E})$;
$\hat{V}_{\vec{B}}$ (\ref{VB}) could be expressed as:
\begin{equation}
\hat{V}_{\vec{B}}=\beta_{S}B_{eff}^{2}-\beta_{T}B_{eff}^{2}\left(\vec{S}\vec{n}_{B}\right)^{2},~
\vec{n}_B=\frac{\vec{B}-\vec{\beta} \times
\vec{E}}{|\vec{B}-\vec{\beta} \times \vec{E}|}.
%\eqno(1.5)
 \label{VB1}
\end{equation}
where $\beta_{S}$ is the scalar magnetic polarizability and
$\beta_{T}$ is the tensor magnetic polarizability of the particle.

$\hat{V}_{{B}}^{nucl}$ describes the effective potential energy of
particle magnetic moment interaction the pseudomagnetic field of
the target \cite{2rot}-\cite{6rot},\cite{birefringence}.

$\hat{V}_E^{nucl}$ describes the effective potential energy of
particle electric moment interaction the pseudoelectric field of
the target \cite{2rot}-\cite{6rot},\cite{birefringence}.

It should be emphasized that $\hat{V}_{{B}}^{nucl}$ and
$\hat{V}_E^{nucl}$ include contributions from strong interactions
as well as those caused by weak interaction violating P (space)
and T (time) invariance.

\section{The equations describing the spin evolution of a particle in a storage ring}

 Let us consider particles moving in a storage ring with low
pressure of residual gas ($10^{-10}$ Torr) and without targets
inside the storage ring.
 In this case we can omit the effects
caused by the interactions $\hat{V}_{{B}}^{nucl}$ and
$\hat{V}_E^{nucl}$.

 Let us consider a particle with $S=1$ (for example, deuteron)
moving in a storage ring.
 According to the above analysis spin
behavior of such a particle can not be described by the
Bargman-Myshel-Telegdy equation.
 The equations for particle spin
motion including contribution from the tensor electric
polarizability were obtained in \cite{birefringence,nastya}.
Considering that deuteron possesses also the tensor magnetic
polarizability and adding the terms caused by it to the equations
obtained in \cite{birefringence,nastya} finally we get:
\begin{eqnarray}
\left\{
\begin{array}{l}
\frac{d\vec{P}}{dt}=
\frac{e}{mc}\left[\vec{P}\times\left\{\left(a+\frac{1}{\gamma}\right)\vec{B}
-a\frac{\gamma}{\gamma+1}\left(\vec{\beta}\cdot\vec{B}\right)\vec{\beta}-
\left(\frac{g}{2}-\frac{\gamma}{\gamma+1}\right)\vec{\beta}\times\vec{E}\right\}\right]+\\
 +
\frac{d}{\hbar}\left[\vec{P}\times\left({\vec{E}}+\vec{\beta}\times\vec{B}\right)\right]
-\frac{2}{3}\frac{\alpha_{T}E^{2}_{eff}}{\hbar}[\vec{n}_{E}\times\vec{n}_{E}^{\prime}]
-\frac{2}{3}\frac{\beta_{T}B^{2}_{eff}}{\hbar}[\vec{n}_{B}\times\vec{n}_{B}^{\prime}],\\
{} \\
\frac{dP_{ik}}{dt}  =
-\left(\varepsilon_{jkr}P_{ij}\Omega_{r}+\varepsilon_{jir}P_{kj}\Omega_{r}\right)
 - \\
-
\frac{3}{2}\frac{\alpha_{T}E^{2}_{eff}}{\hbar}\left([\vec{n}_{E}\times\vec{P}]_{i}n_{E,\,k}
+n_{E,\,i}[\vec{n}_{E}\times\vec{P}]_{k}\right)-\\
 -
\frac{3}{2}\frac{\beta_{T}B^{2}_{eff}}{\hbar}\left([\vec{n}_{B}\times\vec{P}]_{i}n_{B,\,k}
+n_{B,\,i}[\vec{n}_{B}\times\vec{P}]_{k}\right),
\\
\end{array}
\right. \label{50}
\end{eqnarray}
where $m$ is the mass of the particle, $e$ is its charge,
$\vec{P}$ is the spin polarization vector,
$P_{xx}+P_{yy}+P_{zz}=0$,
$\gamma$ is the Lorentz-factor,
 $\vec{\beta}=\vec{v}/c$,
$\vec{v}$ is the particle velocity,
$a=(g-2)/2$, $g$ is the gyromagnetic ratio, $
\vec{E}$ and
 $\vec{B}$ are the electric and magnetic fields in the point of
 particle location, $\vec{E}_{eff}=(\vec{E}+\vec{\beta} \times
\vec{B})$, $\vec{B}_{eff}=(\vec{B}-\vec{\beta} \times \vec{E})$,
$\vec{n}=\vec{k}/k$,
$\vec{n}_{E}=\frac{\vec{E}+\vec{\beta}\times\vec{B}}{|\vec{E}+\vec{\beta}\times\vec{B}|}$,
$\vec{n}_B=\frac{\vec{B}-\vec{\beta} \times
\vec{E}}{|\vec{B}-\vec{\beta} \times \vec{E}|}$,
 $n_{i}^{\prime}=P_{ik}n_{k}$,
$n_{E,\,i}^{\prime}=P_{ik}n_{E,\,k}$,
$n_{Bi}^{\prime}=P_{il}n_{Bl}=P_{i3}$, $\Omega_{r}(d)$ are the
components of the vector $\vec{\Omega}(d)$ ($r=1,2,3$ correspond
to $x,y,z$, respectively).
\begin{eqnarray}
\vec{\Omega}(d) & = &
\frac{e}{mc}\left\{\left(a+\frac{1}{\gamma}\right)\vec{B}
-a\frac{\gamma}{\gamma+1}\left(\vec{\beta}\cdot\vec{B}\right)\vec{\beta}-
\left(\frac{g}{2}-\frac{\gamma}{\gamma+1}\right)\vec{\beta}\times\vec{E}\right\} + \nonumber \\
& + &
\frac{d}{\hbar}\left({\vec{E}}+\vec{\beta}\times\vec{B}\right).
%\eqno(2.24)
\label{2.24}
\end{eqnarray}

\section{Deuteron birefringence effect in electromagnetic field}

When omitting contribution from interaction of the particle EDM
with the electric field $\hat{V}_{EDM}$ we can rewrite the
equations for particle spin motion (\ref{50}) as follows:
\begin{eqnarray}
\frac{d\vec{P}}{dt}=[\vec{P}\times\vec{\Omega}]+\Omega_T[\vec{n}_E\times\vec{n}_E^{\prime}]+
\Omega_T^{\mu}[\vec{n}_B\times\vec{n}_B^{\prime}], \nonumber\\
{} \nonumber\\
\frac{d\vec{P_{ik}}}{dt}=-(\epsilon_{jkr}P_{ij}\Omega_r+\epsilon_{jir}P_{kj}\Omega_r)+
\Omega_T^{\prime}([\vec{n}_E\times\vec{P}]_i
n_{Ek}+n_{Ei}[\vec{n}_E\times\vec{P}]_k) + \nonumber \\
{} \nonumber\\
+\Omega_T^{\prime \mu}([\vec{n}_B\times\vec{P}]_i
n_{Bk}+n_{Bi}[\vec{n}_B\times\vec{P}]_k)\label{BMT} \label{BMT+}
\end{eqnarray}
 where
\begin{eqnarray}
\begin{array}{l}
\vec{\Omega}=\frac{e}{mc}\left[\left(a+\frac{1}{\gamma}\right)\vec{B}
-a\frac{\gamma}{\gamma+1}\left(\vec{\beta}\cdot\vec{B}\right)\vec{\beta}-
\left(\frac{g}{2}-\frac{\gamma}{\gamma+1}\right)\vec{\beta}\times\vec{E}\right], \nonumber \\
{}\nonumber \\
\Omega_T=-\frac{2}{3} \frac{\alpha_T E_{eff}^2}{\hbar} ,
~~\Omega_T^{\prime}=-\frac{3}{2} \frac{\alpha_T E_{eff}^2}{\hbar},
~~\Omega_T^{\prime}=-\frac{2}{3} \Omega_T,\nonumber \\
{}\nonumber \\
 \Omega_T^{\mu}=-\frac{2}{3} \frac{\beta_T
B_{eff}^2}{\hbar} , ~~\Omega_T^{\prime \mu}=-\frac{3}{2}
\frac{\beta_T B_{eff}^2}{\hbar}, ~~\Omega_T^{\prime
\mu}=-\frac{2}{3} \Omega_T^{\mu}.
\end{array}
\end{eqnarray}

Thus presence of the electric and magnetic tensor polarizabilities
makes impossible to describe the spin evolution of a particle in a
 by the Bargman-Myshel-Telegdy equation
\begin{equation}
\frac{d\vec{P}}{dt}=[\vec{P}\times\vec{\Omega}]
\label{BMT}
\end{equation}
but requires considering of the system (\ref{BMT+}).

Let us consider the coordinate system and vectors
$\vec{v},\vec{E}, \vec{B}$ as shown in figure and denote the axes
by $x,y,z$ (or $1,2,3$, respectively).
\begin{figure}[h]
\begin{center}
\begin{picture}(40,40)
\thicklines \put(0,0) {\vector(1,0){40}}
 \put(0,0){\vector(0,1){25}}
\put(-12,12){$\vec{B}$}
  \put(0,0){\vector(0,1){40}}
 \put(0,0){\vector(-3,-4){20}}
 \put(0,0){\vector(-3,-4){12}}
 \put(-6,-20){$\vec{E}$}
 \put(38,2){y}
 \put(4,34){z}
  \put(-22,-20){x}
 \put(0,0){\vector(1,0){25}}
 \put(16,-12){$\vec{v}$}
\end{picture}
\end{center}
\caption{}
\label{Fig.1}
\end{figure}

Suppose that an electric field is absent and the particle initial
polarization coincides with $\vec{v}$ direction, therefore, the
components of the vectors are:
\begin{eqnarray}
\begin{array}{l}
\vec{P}=\left(P_1,P_2,P_3\right), \vec{P}_0=\left(0,P,0\right),\\
  {}  \\
 \vec{n}_E=\left(1,0,0\right), n_{Ei}^{\prime}=P_{il}n_{El}=P_{i1}
   \\
  {} \\
%\end{eqnarray}
 {[} \vec{n}_E\times\vec{n}_{E}^{\prime}{]}_1=0,~{[}\vec{n}_E\times\vec{n}_{E}^{\prime}{]}_2=-P_{31},~
 {[}\vec{n}_E\times\vec{n}_{E}^{\prime}{]}_3=P_2,
 \\
{}  \\
 {[}\vec{P}\times\vec{\Omega} {]}_1=\Omega P_2,~
 {[} \vec{P}\times\vec{\Omega}{]}_2=-\Omega P_1,~
 {[} \vec{P}\times\vec{\Omega}{]}_3=P_2, \\
{}\\
 {[} \vec{n}_E\times\vec{P}{]}_1=0,~
 {[} \vec{n}_E\times\vec{P}{]}_2=-P_3,~
 {[} \vec{n}_E\times\vec{P}{]}_3=P_2,
\end{array}
\label{E}
\end{eqnarray}
\begin{eqnarray}
\begin{array}{l}
\vec{\Omega}=\frac{e}{mc}\left(a+\frac{1}{\gamma}\right)\vec{B}=\left(0,0,\Omega\right),
 \\
  {} \\
 \vec{n}_B=\left(0,0,1\right), n_{Bi}^{\prime}=P_{il}n_{Bl}=P_{i3}
 \\
  {} \\
%\end{eqnarray}
 {[}\vec{n}_B\times\vec{n}_{B}^{\prime}{]}_1=-P_{23},~{[}\vec{n}_B\times\vec{n}_{B}^{\prime}{]}_2=-P_{13},~
 {[} \vec{n}_B\times\vec{n}_{B}^{\prime}{]}_3=0,
\\
{} \\
 %{[} \vec{P}\times\vec{\Omega} {]}_1=\Omega P_2,~
 %{[} \vec{P}\times\vec{\Omega}{]}_2=-\Omega P_1,~
 %{[} \vec{P}\times\vec{\Omega}{]}_3=P_2, \nonumber \\
 {[} \vec{n}_B\times\vec{P}{]}_1=-P_2,~
 {[} \vec{n}_B\times\vec{P}{]}_2=P_1,~
 {[} \vec{n}_B\times\vec{P}{]}_3=0.
 \label{B}
\end{array}
\end{eqnarray}

Substituting (\ref{E},\ref{B}) to the system (\ref{50}) we obtain:
\begin{eqnarray}
\begin{array}{l}
\frac{dP_1}{dt}=\Omega P_2-\Omega_T^{\mu} P_{23},\\
{}  \\
 \frac{dP_2}{dt}=-\Omega P_1+(\Omega_T^{\mu}-\Omega_T)P_{13},\\
 {}  \\
 \frac{dP_3}{dt}=\Omega_T P_{12}
 \end{array}
\label{Pi}
\end{eqnarray}
\begin{eqnarray}
\begin{array}{l}
\frac{dP_{11}}{dt} = 2 \Omega_3 P_{12},  \\
 {}  \\
\frac{dP_{22}}{dt} = -2 \Omega_3 P_{12}, \\
 {}  \\
\frac{dP_{33}}{dt} =0, \\
 \end{array}
\label{sys1}
\end{eqnarray}
\begin{eqnarray}
\begin{array}{l}
\frac{dP_{12}}{dt} = -\Omega \left(
P_{11}-P_{22}\right)-\Omega_T^{\prime} P_3,
 \\
 {}  \\
\frac{dP_{13}}{dt} = \Omega P_{23}+{\Omega}_T^{\prime}P_2-{\Omega}_T^{\prime \mu}P_2,  \\
{}  \\
\frac{dP_{23}}{dt} = -\Omega P_{13}+{\Omega}_T^{\prime \mu}P_1
 \end{array}
\label{p12def}
\end{eqnarray}
remembering that $P_{11}+P_{22}+P_{33}=0$ and $P_{ik}=P_{ki}$,
then getting $P_{33}=const$ from the last equation in
(\ref{sys1})we can conclude that $P_{11}+P_{22}=const$
%Suppose that $P_1$ and $P_2$ rotates with the frequency $\Omega$
%than it can be seen from the above that modulation of
%${\Omega}_T^{\prime \mu}$ components $P_{13}$ and $P_{23}$
%converts to each other.

\subsection{Contribution from the tensor electric polarizability to deuteron spin oscillation}

 From the system (\ref{sys1}) it follows

\begin{eqnarray}
\begin{array}{l}
\frac{d(P_{11}-P_{22})}{dt} = 4\Omega P_{12},
\\
 {}  \\
\frac{d^2P_{12}}{dt^2} = -\Omega
\frac{d(P_{11}-P_{22})}{dt}-{\Omega}_T^{\prime} \frac{dP_3}{dt}=
-(4\Omega^2+\Omega_{T}{\Omega}_T^{\prime})P_{12}.  \\
 \end{array}
\label{p11-p22}
\end{eqnarray}

Thus we have the equation
\begin{equation}
\frac{d^2P_{12}}{dt^2}+ \omega_{12}^2 P_{12}=0
\end{equation}
where $\omega_{12}=\sqrt{4\Omega^2+\Omega_{T}{\Omega}_T^{\prime}}
\approx 2 \Omega$, because $\Omega_{T}{\Omega}_T^{\prime} \ll
\Omega^2$.

The solution for this equation can be found in the form:
\begin{equation}
P_{12}=c_1 \cos{\omega_{12} t}+c_2 \sin{\omega_{12} t} \label{P12}
\end{equation}

Let us find coefficients $c_1$ and $c_2$: when $t=0$ the equation
(\ref{P12}) gives $c_1=P_{12}(0)$. The coefficient $c_2$ can be
found from
\begin{equation}
\frac{d(P_{12})}{dt}(t \rightarrow 0)=\omega_{12} c_2,
\end{equation}
therefore
\begin{equation}
c_2= \frac{1}{\omega_{12}} \frac{d(P_{12})}{dt}(t \rightarrow 0) ,
\end{equation}
From the equation (\ref{p12def})
\begin{equation}
\frac{dP_{12}}{dt} (t \rightarrow 0)= -\Omega \left( P_{11}(t
\rightarrow 0)-P_{22}(t \rightarrow 0)\right), \label{c2}
\end{equation}
that
\begin{equation}
c_2=-\frac{P_{11}-P_{22}}{2}, \label{c2a}
\end{equation}
and
\begin{equation}
P_{12}=P_{12}(0) \cos{\omega_{12} t}-\frac{P_{11}-P_{22}}{2}
\sin{\omega_{12} t} \label{P12final}
\end{equation}

As a result we can write the following equation for the vertical
component of the spin $P_3$:
\begin{equation}
\frac{dP_{3}}{dt}=\Omega_T P_{12}(t)= \Omega_T [ P_{12}(0)\cos{2
\Omega t}-\frac{P_{11}(0)-P_{22}(0)}{2} \sin{2\Omega t} ]
\label{P12final1}
\end{equation}

As it can be seen the vertical component of the spin oscillates
with the frequency $2 \Omega$.

But it should be mentioned that according to the equations
(\ref{50}) interaction of the EDM with an electric field causes
oscillations of the vertical component of the spin with the
frequency $\Omega$. According to the idea \cite{orlov} these
oscillations can be eliminated if the deuteron velocity is
modulated with the frequency $\Omega$:
\begin{equation}
v=v_0+\delta v \sin{(\Omega t + \varphi)} \label{v}
\end{equation}
 As a result $E_{eff}$ depends on $\vec{\beta}=\vec{v}/c$ it also appears
modulated:
\begin{equation}
E_{eff}=E_{eff}^0+\delta E_{eff} \sin{(\Omega t + \varphi)}
\end{equation}
here $\varphi$ is a phase. Therefore,
\begin{equation}
\frac{dP_3}{dt}=\Omega_T P_{12} - d E_{eff} P_2
\end{equation}
as $P_2$ also oscillate with $\Omega$ frequency, then in the
product $E_{eff} P_2$ there non-oscillating terms and $P_3$
linearly grows with time.

It is important that modulation of the velocity $v=v_0+\delta v
\sin{(\Omega t + \varphi)}$ results in oscillation of $E_{eff}^2$
also oscillates with time and appears proportional to $\sin^2
{(\Omega_T + \varphi)}$.
 As a result
\begin{equation}
\frac{dP_3}{dt} \sim \Delta \Omega_T \sin^2 {(\Omega_T t +
\varphi)}[ P_{12}(0)\cos{2 \Omega t}-\frac{P_{11}(0)-P_{22}(0)}{2}
\sin{2\Omega t} ]
\end{equation}
 i.e.
\begin{equation}
\frac{dP_3}{dt} \sim - \frac{1}{2} \Delta \Omega_T \cos {(2 \Omega
t+ 2\varphi)}[ P_{12}(0)\cos{2 \Omega
t}-\frac{P_{11}(0)-P_{22}(0)}{2} \sin{2\Omega t} ] \label{28}
\end{equation}

According to (\ref{28}) if the phase $\varphi=0$ then the
contribution to the linear growth of $P_3$ is provided by the term
$P_{12}(0)$.
 If $\varphi=\pi/4$ then linear growth is due to the second term
proportional to $(P_{11}(0)-P_{22}(0))$.

Measurement of these contribution provides to measure the tensor
electric polarizability.

According to the evaluations \cite{polarizability} $\alpha_T \sim
10^{-40}$ cm$^3$  for the field $E_{eff}=B \sim 10^4$ gauss,
therefore the frequency $ \Omega_T \sim 10^{-5}$ sec$^{-1}$. When
considering modulation we should estimate $\Delta \Omega_T \sim
\Omega_T {(\frac{\delta}{v_0})}^2$, then  suppose
${(\frac{\delta}{v_0})}^2 \sim 10^{-2} - 10^{-3}$ we obtain
$\Delta \Omega_T \sim 10^{-7} - 10^{-8}$ sec$^{-1}$.

\subsection{Contribution from the tensor magnetic polarizability to deuteron spin oscillation}

Let us consider now contributions caused by the tensor magnetic
polarizability $\beta_T$.
% According to the calculations \cite{polarizability} $\alpha_T<\beta_T$.
Let we omit the terms proportional to the tensor electric
polarizability in the system (\ref{sys1}):
\begin{eqnarray}
\begin{array}{l}
\frac{dP_1}{dt}=\Omega P_2-\Omega_T^{\mu} P_{23},\\
{}  \\
 \frac{dP_2}{dt}=-\Omega P_1+\Omega_T^{\mu} P_{13},\\
 {}  \\
 \frac{dP_{13}}{dt} = \Omega P_{23}-{\Omega}_T^{\prime \mu}P_2,  \\
{}  \\
\frac{dP_{23}}{dt} = -\Omega P_{13}+{\Omega}_T^{\prime
\mu}P_1
\end{array} \label{sys2}
\end{eqnarray}

Introducing new variables $P_{+}=P_1+iP_2$ and
$G_{+}=P_{13}+iP_{23}$ and recomposing equations (\ref{sys2}) to
determine $P_{+}$ and $G_{+}$ we obtain:
\begin{eqnarray}
\begin{array}{l}
\frac{dP_+}{dt}=-i\Omega P_+ + i\Omega_T^{\mu} G_+,\\
{} \nonumber \\
 \frac{dG_+}{dt}=-i\Omega G_+ + i\Omega_T^{\prime \mu} P_+,\\
\end{array}
\label{sys22}
\end{eqnarray}

or

\begin{eqnarray}
\begin{array}{l}
i\frac{dP_+}{dt}=\Omega P_+ - \Omega_T^{\mu} G_+,\\
{} \nonumber \\
i \frac{dG_+}{dt}=\Omega G_+ - \Omega_T^{\prime \mu} P_+,\\
\end{array}
\label{sys3}
\end{eqnarray}

Let us search $P_+,G_+ \sim e^{i\omega t}$ then (\ref{sys3})
transforms as follows:

\begin{eqnarray}
\begin{array}{l}
\omega \tilde{P}_+=\Omega \tilde{P}_+ - \Omega_T^{\mu} \tilde{G}_+,\\
{} \nonumber \\
\omega \tilde{G}_+=\Omega \tilde{G}_+ - \Omega_T^{\prime \mu}
\tilde{P}_+.
\end{array}
\label{sys33}
\end{eqnarray}

The solution of this system can be easily find:
\begin{eqnarray}
\begin{array}{l}
(\omega -\Omega)^2- \Omega_T^{\mu} \Omega_T^{\prime \mu}=0\\
\end{array}
\label{sys34}
\end{eqnarray}
 that finally gives
 \begin{equation}
\omega_{1,2}=\Omega \pm \sqrt{\Omega_T^{\mu} \Omega_T^{\prime
\mu}}
\end{equation}

Rewriting the solution
 \begin{equation}
P_+(t)=c_1 e^{-i\omega_1 t}+c_2 e^{-i\omega_2 t}= |c_1|
e^{-i(\omega_1 t+\delta_1)}+|c_2| e^{-i(\omega_2 t+\delta_2)}
\label{P+}
\end{equation}
Therefore,
 \begin{equation}
P_1(t)=|c_1| \cos (\omega_1 t+\delta_1)+|c_2| \cos (\omega_2
t+\delta_2) \label{reP+}
\end{equation}

 This means that spin rotates with two frequencies $\omega_1$ and
 $\omega_2$ instead of $\Omega$ and, therefore, experiences beating with the frequency
$\Delta \omega=\omega_1-\omega_2=2 \sqrt{\Omega_T^{\mu}
\Omega_T^{\prime \mu}}=\frac{\beta_T B_{eff}^2}{\hbar} $.

According to the evaluation \cite{polarizability} the tensor
magnetic polarizability $\beta_T \sim 2 \cdot 10^{-40}$, therefore
for the beating frequency $\Delta \omega \sim 10^{-5}$ in the
field $B \sim 10^4$ gauss.

Measurement of the frequency of this beating makes possible to
measure the tensor magnetic polarizability of the deuteron
(nuclei).

Thus, due to the presence of tensor magnetic polarizability the
the horizontal component of spin rotates around $\vec{B}$ with two
frequencies $\omega_1,~\omega_2$ instead of expected rotation with
the frequency $\Omega$. The resulting motion of the spin is
beating: $P_1(t) \sim \cos \Omega t \sin \Delta \omega t$.

This is the reason for the component $P_3$ caused by the EDM to
experience the similar beating. Therefore, particle velocity
modulation with the frequency $\Omega$ ($v=v_0+\delta v
\sin{(\Omega t + \varphi)}$) provides for eliminating oscillation
with $\Omega$ frequency, but oscillations with the frequency
$\Delta \omega$ rest.

\section{Spin rotation of proton (deuteron, antiproton) in a storage ring with a polarized target
and paramagnetic resonance in the nuclear pseudoelectric and
pseudomagnetic fields }

Another class of experiments deals with the use of polarized
targets. Preparing such experiment one should remember that
density of polarized gas target is lower than nonpolarized that
and for example for COSY density of polarized target is $j =
10^{14} cm^{-2}$.

In 1964 it was shown \cite{1rot} that while slow neutrons are
propagating through the target with polarized nuclei a new effect
of nucleon spin precession occurred. It is stipulated by the fact
that in a polarized target the neutrons are characterized by two
refraction indices ($N_{\uparrow \uparrow }$ for neutrons with the
spin parallel to the target polarization vector and $N_{\uparrow
\downarrow }$ for neutrons with the opposite spin orientation ,
$N_{\uparrow \uparrow }\neq N_{\uparrow \downarrow })$. According
to the \cite{2rot}, in the target with polarized nuclei there is a
nuclear pseudomagnetic field and the interaction of an incident
neutron with this field results in neutron spin rotation. The
results obtained in \cite{1rot}, initiated experiments which
proved the existence of this effect \cite{9rot}-\cite{11}.

The effective potential energy of a particle in the pseudomagnetic
nuclear field $\vec{G}$ of matter can be written as:

\begin{equation}
\hat{V}_{B}^{nucl}=-\vec{\mu} \vec{G},
 \label{Vnucl}
\end{equation}
where $\vec{\mu}$ is the magnetic moment of the particle and
$\vec{G}$ can be expressed as \cite{2rot}-\cite{6rot}
\begin{eqnarray}
\begin{array}{c}
\vec{G}=\vec{G}_s+\vec{G}_w, \\
{} \\
 \vec{G}_s=\frac{2 \pi \hbar^2}{\mu m}\rho [A_1 \langle
\vec{J}
\rangle + A_2 \vec{n} (\vec{n}\langle \vec{J} \rangle)+...], \\
{} \\
 \vec{G}_w=\frac{2 \pi \hbar^2}{\mu m}\rho [b \vec{n} + b_1
[\langle \vec{J} \rangle \times \vec{n}]+b_2 \vec{n}_1
+b_3\vec{n}(\vec{n}\vec{n}_1)+b_5 [\vec{n} \times \vec{n}_1]+...]
\end{array}
 \label{Gnucl}
\end{eqnarray}
where $\vec{n}=\vec{v}/v$, $\vec{J}$ is the spin of nuclei of
matter, $\langle \vec{J} \rangle=\textrm{Sp} \rho_{nucl} \vec{J}$
is the average value of nuclear spin, $\vec{n}_1$ has the
components $n_{1j}=\langle Q_{ij} \rangle n_j$, where $\langle
Q_{ij} \rangle=\textrm{Sp} \rho_{nucl} Q_{ij}$ is the polarization
tensor
\begin{equation}
Q_{ij}=\frac{1}{2J(2J-1)}\left\{ J_i J_j + J_j J_i -\frac{2}{3} J
(J+1) \delta_{ij}\right\},
 \label{tensor polarization}
\end{equation}

It is easy to see that interaction (\ref{Vnucl}) looks like the
interaction of a magnetic moment with a magnetic field, thus the
field $\vec{G}$ contributes to the change of the particle
polarization similar a magnetic field does. It should be
especially mentioned that $\hat{V}_{B}^{nucl}$ contains both the
real part, which is responsible for spin rotation, and imaginary
part, which contributes to spin dichroism (i.e. beam absorption
dependence on spin orientation). The detailed analysis of the
effects caused by the nuclear pseudoelectric field was done in
\cite{birefringence}.

Interaction with the field $ \vec{G}=\vec{G}_s+\vec{G}_w$ contains
two summands: the first $\vec{G}_s$ corresponds to the strong
interaction, which is T,P-even, while the second $\vec{G}_w$
describes spin rotation by the weak interaction, which has both
T,P-odd (the term containing the constant $b_1$) and T-odd, P-even
(the term containing the constant $b_5$) terms.

If either vector or tensor polarization of a target rotates then
the effects provided by $\vec{G}_s$, $\vec{G}_w$ periodically
depend on time i.e. equation (\ref{1}) converts to:
\begin{equation}
i\hbar\frac{\partial\Psi(t)}{\partial
t}=\left(\hat{H}_{0}+\hat{V}_{EDM}+\hat{V}_{\vec{E}}+\hat{V}_{\vec{B}}+\hat{V}_E^{nucl}(t)+
\hat{V}_{{B}}^{nucl}(t)\right)\Psi(t). \label{1(t)}
\end{equation}
This equation coincides with the well-known equation for the
paramagnetic resonance. Really, if we have the strong field
orthogonal to the weak one (in this case $\vec{B} \perp \vec{G}$)
and $\vec{G}$ either rotates or oscillates with the frequency
corresponding to the splitting, caused by the field $\vec{B}$, the
resonance occurs. In our case this leads to the conversion of
horizontal spin component to the vertical one with the frequency
determined by the frequency of spin precession in the field
$\vec{G}$. Thus we can measure all the constants containing in
$\vec{G}_s$ and $\vec{G}_w$: constants $A_i$ give the
spin-dependent part of elastic coherent forward scattering
amplitude of proton (deuteron, antiproton) that is important for
the projects at GSI and COSY; amplitudes $b_i$ provides to measure
the constants of T-,P-odd interactions.

 First of all we should pay attention to the effects caused by the T-odd
 nucleon-nucleon
 interaction of protons (antoprotons) and deuterons with polarized nuclei and, in particular, interaction
 described by $V_{P,T} \sim \vec{S} \left[ \vec{p}_N \times \vec{n}
 \right]$, where $\vec{P}_N(t)$ is the polarization vector of
 target.
%\\\\\\\\\\\\\\\\\\\\\\\\\\\\\\\\\\\\\\\\\
  The interaction $V_{P,T}$ leads to the spin rotation around the
  axis  determined by the unit vector $\vec{n}_T$ parallel to the
  vector $\left[ \vec{P}_N(t) \times \vec{n}
 \right]$.
 Spin dichroism also appears with respect to this vector
 $\vec{n}_T$ i.e. a proton (deuteron) beam with the spin parallel
 to $\vec{n}_T$ has the absorption cross-section different from
 the absorption cross-section for a proton (deuteron) beam with the opposite
 spin direction.
%\\\\\\\\\\\\\\\\\\\\\\\\\\\\\\\\\\\\\\\\\

 P-even T-odd spin rotation, oscillation and dichroism of deuterons (nuclei with $S \ge 1$)
  caused by the interaction either $V_T \sim (\vec{S} [ \vec{P}_N(t) \times \vec{n}
  ])(\vec{S}\vec{n})$ could be observed \cite{birefringence};
  P-even T-odd spin rotation and dichroism for a proton, deuteron (nucleus with the spin $S \ge 1/2$)
  $V^{\prime}_T \sim  b_5 [\vec{n} \times \vec{n}_1(t)]$
  could be observed
  \cite{4} in paramagnetic resonance conditions, too.

\section{Conclusion}

In the present paper the equations for the spin evolution of a
particle in a storage ring are analyzed considering contributions
from the tensor electric and magnetic polarizabilities of the
particle. Study of spin rotation and birefringence effect for a
particle in a high energy storage ring provides for measurement as
the real part of the coherent elastic zero-angle scattering
amplitude as well as tensor electric and magnetic
polarizabilities.

We proposed the method for measurement the real part of the
elastic coherent zero-angle scattering amplitude of particles and
nuclei in a storage ring by the paramagnetic resonance in the
periodical in time nuclear pseudoelectric and pseudomagnetic
fields.

% ---------------------------------- from wind - literature correction

\end{document}